\documentclass[12pt]{article}
\usepackage{epsf}
\usepackage{amsfonts} 
\usepackage{amsmath} 
\usepackage[latin2]{inputenc} 
\usepackage[T1]{fontenc} 
\usepackage{epsfig} 
 
\newlength{\dinwidth}                          
\newlength{\dinmargin} 
\DeclareMathAlphabet{\scr}{U}{rsfs}{m}{n}  
\begin{document} 
\pagestyle{empty}  
\begin{flushright} 
IFT-2003-07
\end{flushright}  
\vskip 2cm  
\begin{center}  
{\Huge  
Twisted supergravity and untwisted super--bigravity}
\vspace*{5mm} \vspace*{1cm}   
\end{center}  
\vspace*{5mm} \noindent  
\vskip 0.5cm 
\centerline{\bf Zygmunt Lalak and Rados\l aw Matyszkiewicz
}  
\vskip 1cm   
\centerline{\em Institute of Theoretical Physics}  
\centerline{\em University of Warsaw, Poland}  
\vskip3cm
\centerline{\bf Abstract}  
\vskip 0.5cm
We have extended previous analysis of the bulk/brane 
supersymmetrizations involving non-zero brane mass terms of bulk fermions 
(gravitini) and twisting of boundary conditions.  
We have constructed  new brane/bulk models that may be relevant for 
realistic model building. In particular, we have built a model with the 
Randall--Sundrum bosonic sector, orthogonal projection operators on the branes in the fermionic sector, and an unbroken $N=1$ supersymmetry. 
We have also constructed 5d super--bigravity with static 
vacuum and unbroken $N=1$ supersymmetry, which may be viewed as a 
deconstruction of 5d supergravity.    
\vskip3cm  
\begin{flushleft}March 2003\end{flushleft}
\newpage  
\pagestyle{plain}   
\section{Introduction}
Higher--dimensional theories with branes offer an intriguing possibility 
of breaking symmetries through imposing, at the
branes, nontrivial boundary conditions on fields transforming in
various representations of the subgroups of the original symmetry group. 
Subsequently, when the breaking is explicit at the distant brane, 
locality in extra--dimension implies that the result of that breaking seen 
at the observable brane becomes conveniently suppressed. 
At the same time one believes that the ultimate solution to the hierarchy problem is related to space-time supersymmetry, and one wants to retain the gravitational sector in the model, in the hope to partially unify gravity with other 
forces and to explain the Planck scale/weak scale hierarchy.
Hence, since supersymmetry is actually a space-time symmetry, the natural 
arena to study symmetry breaking in brane worlds is the higher--dimensional supergravity with branes. Such theories have been constructed in five dimensions
in \cite{Altendorfer:2000rr}--\cite{Brax:2002vs}, and we shall 
use them as the framework of our discussion. 
In fact, we shall concentrate on the pure supergravity case, and the symmetry 
broken by boundary conditions will be the supersymmetry itself. 
The extensive analysis of such a breakdown in the particular case of 
the FLP supergravity \cite{Falkowski:2000er}, coupled to a bulk hypermultiplet, has been 
presented in \cite{Lalak:2002kx}. Here we extend that analysis to the 
bulk/brane 
supersymmetrizations involving non-zero brane mass terms of bulk fermions 
(gravitini). Preliminary analysis and the results of \cite{Lalak:2002kx} 
imply that 
the bulk fields from the hypermultiplet will behave analogously to gravitini. 
Bulk moduli shall play an important role in transmission of supersymmetry breakdown to the matter sectors, however we postpone the analysis of such a 
matter/moduli/gravity system to a future publication. Our results confirm 
these of the ref. \cite{Bagger:2002rw} at points where the papers overlap. 
  
\section{General construction}

To begin with, let us summarize the construction of the general brane--bulk
supergravity Lagrangian using the notation, and in the spirit of \cite{Falkowski:2000er},\cite{Brax:2001xf}. The simple N=2 d=5 supergravity multiplet contains metric tensor (represented by 
the vielbein $e^m_\alpha$), two gravitini $\Psi^A_\alpha$ and one vector field $A_\alpha$ -- the graviphoton. We shall consider gauging of a $U(1)$ subgroup of the global $SU(2)_R$ symmetry of the 5d Lagrangian. In general, coupling of bulk fields to branes turns out to be related to the gauging, and the bulk-brane couplings will preserve only a subgroup of the $SU(2)_R$. 
Purely gravitational 5d action describing such a setup reads $
        S=\int_{M_5}\ e_5 \ {\cal L}_{grav}\ , 
$        where
        \begin{eqnarray} 
        &{\cal L}_{grav}=&\frac{1}{2}R-\frac{3}{4}{\cal F}_{\alpha\beta}{\cal F}^{\alpha\beta}-\frac{1}{2\sqrt{2}}A_\alpha{\cal F}_{\beta\gamma}{\cal F}_{\delta\epsilon}\epsilon^{\alpha\beta\gamma\delta\epsilon}\nonumber\\&&-\frac{1}{2}\bar{\Psi}^A_\alpha\gamma^{\alpha\beta\gamma}D_\beta\Psi_{\gamma A}+\frac{3{\rm i}}{8\sqrt{2}}\left(\bar{\Psi}^A_\gamma\gamma^{\alpha\beta\gamma\delta}\Psi_{\delta A}+2\bar{\Psi}^{\alpha A}\Psi_{A}^{\beta}\right){\cal F}_{\alpha\beta}\nonumber\\&&-\frac{{\rm i}}{\sqrt{2}} {\cal P}_{AB}\bar{\Psi}^A_\alpha\gamma^{\alpha\beta}\Psi_{\beta}^B-\frac{8}{3} {\rm Tr}({\cal P}^{2})\ .
        \end{eqnarray} 
        Covariant derivative contains both gravitational and gauge connections:
        \begin{equation} 
        D_\alpha\Psi_\beta^A=\nabla_\alpha\Psi_\beta^A+ A_\alpha{\cal P}^A_{\;B}\Psi_\beta^B\ ,
        \end{equation} 
where ${\cal P}={\cal P}_a\,{\rm i}\sigma^a$ is the gauge prepotential. The pair of gravitini satisfies symplectic Majorana\- condition
 $\bar{\Psi}^A\equiv\Psi_A^\dagger\gamma_0=(\epsilon^{AB}\Psi_B)^TC$ where $C$ is the charge conjugation matrix and $\epsilon^{AB}$ is antisymmetric $SU(2)_R$ metric (we use the convention $\epsilon_{12}=\epsilon^{12}=1$). 
Supersymmetry transformations are also modified by the gauging
        \begin{eqnarray} 
        &&\delta e^m_\alpha=\frac{1}{2}\bar{\eta}^A\gamma^m\Psi_{\alpha A},\;\;\delta A_\alpha=\frac{{\rm i}}{2\sqrt{2}}\Psi_{\alpha}^A\eta_A,\\&&\delta\Psi_{\alpha}^A=D_\alpha\eta^A- \frac{{\rm i}}{4\sqrt{2}}\left(\gamma_\alpha^{\;\beta\gamma}-4\delta_\alpha^{\;\beta}\gamma^\gamma\right){\cal F}_{\beta\gamma}\eta^A+\frac{\sqrt{2}{\rm i}}{3}{\cal P}^{AB}\gamma_\alpha\eta_B\ .
        \end{eqnarray} 

In the most general case we can define  ${\bf Z}_2$ action on the gravitino sector as:
\begin{eqnarray} \label{gbcond}
        &\Psi^A_\mu(-y)=\gamma_5(Q_0)^A_{\;B}\Psi^B_\mu(y)\ ,\quad\Psi^A_5(-y)=-\gamma_5(Q_0)^A_{\;B}\Psi^B_5(y)\ ,&\nonumber\\&\Psi^A_\mu(\pi r_c-y)=\gamma_5(Q_\pi)^A_{\;B}\Psi^B_\mu(\pi r_c+y)\ ,\quad\Psi^A_5(\pi r_c-y)=-\gamma_5(Q_\pi)^A_{\;B}\Psi^B_5(\pi r_c+y),&
\end{eqnarray}
and the parameters $\eta^A$ of the supersymmetry transformations obey the same boundary conditions as the 4d components of gravitini\footnote{\label{foot1}We assume that all {\em even} fields are {\em smooth},  except $\Psi_5$ which we take to be {\em smooth} function multiplied by $\epsilon^{2}(y)$. The {\em odd} bosonic fields are also assumed to be {\em smooth}, however {\em odd} gravitini are allowed to have discontinuities proportional to the epsilon functions centered at the branes, e.g.: 
$\Psi_-=\epsilon(y)\bar{\Psi}_-$, where $\bar{\Psi}_-$ is smooth.
	  In our calculations we use the formula $\epsilon^{2}(y)\delta(y)=\frac{1}{3}\delta(y)$, when integrated over a fixed point. In the bulk we use  $\epsilon^{2}(y)=1$. In addition we introduce the distribution $\epsilon^{-1}(y)$ in the following way: 
       $h(y)=\epsilon^{-1}(y)f(y) \Longleftrightarrow f(y)=\epsilon(y)h(y)$.
       One can easily check that derivatives of the step-function-type distributions obey
       $\left(\epsilon^{-1}(y)\right)'=-\epsilon^{-2}(y)\delta(y)$, 
     and everywhere $\epsilon^{-1}\epsilon=1$ (even at fixed points). The parameters of the 
     supersymmetry transformations follow the same rules as gravitini.}.
The symplectic Majorana condition 
and the normalization $(Q_{0,\pi})^2=1$ imply $Q_{0,\pi}=(q_{0,\pi})_a \sigma^a$, where $(q_{0,\pi})_a$ are real parameters \cite{Bergshoeff:2000zn}.

We would like to gauge a $U(1)$ subgroup of the global $SU(2)$, to allow for 
a potential energy in the bulk, and, hence, to obtain a nontrivial 
warp factor.  
In the general case \cite{Brax:2001xf} we can choose the prepotential 
of the form
\begin{equation} 
\label{prep}
        P = g_R \epsilon(y) R + g_{S} S,
\end{equation}
where $R=r_a\, {\rm i} \sigma^a$ and  $S=s_a\, {\rm i}\sigma^a$ commutes and anticommutes with $Q_{0,\pi}$, respectively.

Let us explore supersymmetry algebra locally near the point $y=0$. Non--vanishing part of the Lagrangian variation includes terms proportional to $\delta(y)$:
\begin{equation}  \label{initvarbrane}
\delta  (e_5{\cal L}_{grav}) \supset -e_4\delta(y){\rm i}\sqrt{8}g_R \bar{\Psi}_\mu^A\gamma^\mu\gamma^5 R_{A}^{\;B}\eta_B \ .
\end{equation} 
   To cancel above variation, we need to add to the initial Lagrangian 
brane tension and/or gravitini mass term on the brane:
  \begin{equation} 
   e_5 {\cal L}_{grav}\longrightarrow  e_5 {\cal L}_{grav}- e_4\delta(y)\lambda_0- e_4\delta(y)\bar{\Psi}_\mu^A\gamma^{\mu\nu}(M+\gamma_5\bar{M})_{A}^{\;B} \Psi_{\nu B} \ ,
   \end{equation} 
  where $M_{AB}$ and $\bar{M}_{AB}$ are symmetric. Associated contributions to the Lagrangian variation read:
  \begin{eqnarray} \label{addbranevar}
  \delta  (e_5{\cal L}_{grav}) \supset&& e_4\delta(y)\frac{1}{2}\lambda_0\bar{\Psi}_\mu^A\gamma^\mu \eta_A\nonumber\\&&+e_4\delta(y)\epsilon(y)\frac{\lambda_{0}}{2}\bar{\Psi}_\mu^A\gamma^{\mu} (\gamma_5M-\bar{M})_A^{\;B} \eta_B \nonumber\\&&-e_4\delta(y){\rm i}\sqrt{8}\bar{\Psi}_\mu^A\gamma^\mu (M-\gamma_5\bar{M})_{A}^{\;B}P_{B}^{\;C}\eta_C\\&&+ e_4\delta(y)2\bar{\Psi}_\mu^A (M+\gamma_5\bar{M})_A^{\;B} \left(\frac{3{\rm i}}{2\sqrt{2}}\gamma^5F^\mu_{\;\;5}\eta_B-\gamma^{\mu\nu}\partial_\nu\eta_B\right) \nonumber \ .
  \end{eqnarray} 
 Now one is able to cancel the variation (\ref{initvarbrane}). However, we have produced an additional uncanceled term -- the last one in 
(\ref{addbranevar}). To solve this problem, let us redefine the 
$\Psi^A_5$ variation:
  \begin{equation} 
  \delta\Psi^A_5 \longrightarrow\delta\Psi^A_5-\delta(y)(\omega+\gamma_5\bar{\omega})^{A}_{\;B}\gamma_5\eta^B\ .
  \end{equation} 
As the result, the new brane contributions to the Lagrangian variation read:
  \begin{eqnarray} \label{addbranesupvar}
   \delta  (e_5{\cal L}_{grav}) \supset && -e_4\delta(y)\epsilon(y)\frac{\lambda_{0}}{4}\bar{\Psi}_\mu^A\gamma^{\mu}(\gamma_5\omega+\bar{\omega})_{A}^{\;B} \eta_B\nonumber\\&& +e_4\delta(y){\rm i}\sqrt{2}\bar{\Psi}_\mu^A\gamma^\mu P_{A}^{\;B}(\omega+\gamma_5\bar{\omega})_{B}^{\;C} \eta_C \\&& +e_4\delta(y)\bar{\Psi}_\mu^A(\omega+\gamma_5\bar{\omega})_{A}^{\;B} \left(\frac{3{\rm i}}{2\sqrt{2}}\gamma^5F^\mu_{\;\;5}\eta_B-\gamma^{\mu\nu}\partial_\nu\eta_B\right)\nonumber\ .  
  \end{eqnarray} 
Notice that $P_{AB}$ is symmetric. 
The vanishing of the sum of the variations (\ref{initvarbrane}),(\ref{addbranevar}),(\ref{addbranesupvar}) leads to the following constraints:
  \begin{eqnarray} \label{rownanienaznikanie}
    0=&\delta(y)\bar{\Psi}_\mu^A\gamma^\mu\bigg[&(M-\gamma_5\bar{M})_{A}^{\;B}P_{B}^{\;C}-\frac{1}{2}P_{A}^{\;B}(\omega+\gamma_5\bar{\omega})_{B}^{\;C}+g_R\gamma_5 R_{A}^{\;C}+\frac{{\rm i}}{4\sqrt{2}}\lambda_0\delta_{A}^{\;C}\nonumber\\&&+\frac{{\rm i}}{4\sqrt{2}}\lambda_0\epsilon(y)\left(
\gamma_5M_{A}^{\;C}-\bar{M}_{A}^{\;C}-\frac{1}{2}\gamma_5\omega_{A}^{\;C}-\frac{1}{2}\bar{\omega}_{A}^{\;C}\right)\bigg]\eta_C
  \end{eqnarray} 
  and
  \begin{equation} \label{omegamasa}
  0=\delta(y)\bar{\Psi}_\mu^A\left[2(M+\gamma_5\bar{M})_{A}^{\;B}+(\omega+\gamma_5\bar{\omega})_{A}^{\;B}\right]\left(\frac{3{\rm i}}{2\sqrt{2}}\gamma^5F^\mu_{\;\;5}\eta_B-\gamma^{\mu\nu}\partial_\nu\eta_B\right)\ .
  \end{equation} 

On the other hand the gravitini equation of motion and supersymmetry transformations imply boundary condition\footnote{The cancellation of terms proportional to the delta functions in the gravitini equations of motion imply the following boundary conditions: $ \epsilon^{-1}(y) \delta(y) (\gamma_5\delta+Q_{0})_A^{\;B}(\Psi_\mu)_B=\delta(y)2(M+\gamma_{5}\bar{M})_A^{\;B}(\Psi_\mu)_B$. On the other hand, odd bosonic fields vanish on the branes, and constraints $\delta_\eta e^{\hat{5}}_\mu=\delta_\eta e_5^a=\delta_\eta A_\mu=0$ imply boundary conditions for the parameters of the supersymmetry transformations $\eta_A$.}:
    \begin{equation} \label{warunkibrzeg}
  \epsilon^{-1}(y) \delta(y) (\gamma_5\delta+Q_{0})_A^{\;B}\eta_B=\delta(y)2(M+\gamma_{5}\bar{M})_A^{\;B}\eta_B
    \end{equation} 
    and the equation (\ref{omegamasa}) takes the form:
 	\begin{eqnarray} \label{omegamasa2}
    0=\delta(y)\bar{\Psi}_\mu^A\left[ \epsilon^{-1}(y) (\gamma_5\delta+Q_{0})_A^{\;B}+(\omega+\gamma_5\bar{\omega})_{A}^{\;B}\right]\left(\frac{3{\rm i}}{2\sqrt{2}}\gamma^5F^\mu_{\;\;5}\eta_B-\gamma^{\mu\nu}\partial_\nu\eta_B\right)\ .
	 \end{eqnarray} 
  
 To complete our discussion, let us check the supersymmetry algebra. 
One can show that the commutators vanish except the following term 
proportional to the delta function \cite{Bagger:2002rw}:
\begin{equation} \label{comutator}
[\delta_\xi,\delta_\eta]e^\mu_5\supset\epsilon^{-1}(y)\delta(y)\bar{\eta}^A\gamma^\mu\left(\delta+\gamma_5Q_{0}\right)_A^{\;B}\xi_B+\delta(y)\bar{\eta}^A\gamma^\mu\left(\gamma_{5}\omega+\bar{\omega}\right)_A^{\;B}\xi_B\ .
\end{equation} 

We can cancel this term by choosing
	\begin{equation} 
	\omega_A^{\;B}=-\epsilon^{-1}(y)(Q_{0})_A^{\;B}\ ,\quad\bar{\omega}_A^{\;B}=-\epsilon^{-1}(y)\delta_A^{\;B}\ ,
	\end{equation} 
then (\ref{omegamasa2}) is simply satisfied and (\ref{rownanienaznikanie}) takes the form:
 \begin{eqnarray} \label{rownanienaznikanie2}
    0=&\delta(y)\bar{\Psi}_\mu^A\gamma^\mu\bigg[&(M-\gamma_5\bar{M})_{A}^{\;B}P_{B}^{\;C}+\frac{1}{2}\epsilon^{-1}(y)P_{A}^{\;B}(Q_0+\gamma_5\delta)_{B}^{\;C}+g_R\gamma_5 R_{A}^{\;C}\nonumber\\&&+\frac{{\rm i}}{4\sqrt{2}}\lambda_0\delta_{A}^{\;C}+\frac{{\rm i}}{4\sqrt{2}}\lambda_0\epsilon(y)\left(\gamma_5M_{A}^{\;C}-\bar{M}_{A}^{\;C}\right)\nonumber\\&&+\frac{{\rm i}}{4\sqrt{2}}\lambda_0\left(\frac{1}{2}\gamma_5(Q_{0})_{A}^{\;C}+\frac{1}{2}\delta_{A}^{\;C}\right)\bigg]\eta_C\ .
  \end{eqnarray} 

We can repeat the same construction for the second brane at the $y=\pi r_c$. The analogue of the equation (\ref{rownanienaznikanie2}) is:
 \begin{eqnarray} \label{rownanienaznikaniepi}
    0=&\delta(y-\pi r_{c})\bar{\Psi}_\mu^A\gamma^\mu\bigg[&(M_{\pi}-\gamma_5\bar{M}_{\pi})_{A}^{\;B}P_{B}^{\;C}-\frac{1}{2}\epsilon^{-1}(y)P_{A}^{\;B}(Q_\pi+\gamma_5\delta)_{B}^{\;C}-g_R\gamma_5 R_{A}^{\;C}\nonumber\\&&+\frac{{\rm i}}{4\sqrt{2}}\lambda_\pi\delta_{A}^{\;C}-\frac{{\rm i}}{4\sqrt{2}}\lambda_\pi\epsilon(y)\left(\gamma_5(M_{\pi})_{A}^{\;C}-(\bar{M}_{\pi})_{A}^{\;C}\right)\nonumber\\&&+\frac{{\rm i}}{4\sqrt{2}}\lambda_\pi\left(\frac{1}{2}\gamma_5(Q_{\pi})_{A}^{\;C}+\frac{1}{2}\delta_{A}^{\;C}\right)\bigg]\eta_C\ ,
  \end{eqnarray}   
	where we choose
	\begin{equation} \label{omegapi}
	(\omega_{\pi})_A^{\;B}=\epsilon^{-1}(y)(Q_{\pi})_A^{\;B}\ ,\quad(\bar{\omega}_{\pi})_A^{\;B}=\epsilon^{-1}(y)\delta_A^{\;B}\ ,
	\end{equation} 
	and boundary condition reads:
	  \begin{equation} \label{warunkibrzegpi}
  \epsilon^{-1}(y) \delta(y-\pi r_{c}) (\gamma_5\delta+Q_{\pi})_A^{\;B}\eta_B=-\delta(y-\pi r_{c})2(M_{\pi}+\gamma_{5}\bar{M}_{\pi})_A^{\;B}\eta_B\ .
    \end{equation} 
The equations (\ref{rownanienaznikanie2}) and (\ref{rownanienaznikaniepi}) 
provide 
the relations, one for each brane, between parameters of the
boundary Lagrangian and the prepotential (whose square determines  the vacuum 
energy in the bulk). These are the conditions analogous, although somewhat weaker, to the strict relation between the brane tensions and the bulk cosmological constant found in the original FLP scenario \cite{Falkowski:2000er}.
We should note at this point that there exists an additional condition that allows to close the susy algebra: one needs to assume that the elementary bosonic fields that are odd with respect to a given brane are also continuous - hence vanishing - on that brane. This is needed in all consistent approaches to bulk/brane supergravity known so far. However, this requirement cannot be imposed on 
non-elementary bosonic fields, like gauge field strengths or spin connections.
On one hand extending the conditions that all odd bosonic fields are continuous on the branes would help to cancel a number of supersymmetry variations 
and to simplify the constraints (\ref{rownanienaznikanie2}), (\ref{rownanienaznikaniepi}). On the other hand such a strong assumption wouldn't allow for a 
supersymmetrization of a wide class of nontrivial models - for instance 
in the Randall--Sundrum model, \cite{Randall:1999ee}, the spin connection is discontinuous 
on the branes (proportional to the step function). 
 
\section{Twisting the Randall--Sundrum model}

In the forthcoming sections we shall apply the formalism developed above 
to the construction of supergravity models with nontrivial boundary 
conditions imposed on the bulk fields. 

\subsection{Randall-Sundrum model without mass terms on branes \cite{Falkowski:2000er}}
To start with let us choose the prepotential in the form: 
$P_A^{\;B}=g\epsilon(y){\rm i}(\sigma_3)_A^{\;B}$. We define ${\bf Z}_2$ action on the gravitini by $(Q_0)_A^{\;B}=(Q_\pi)_A^{\;B}=(\sigma_3)_A^{\;B}$. We do not put any gravitini mass terms on the branes. 
Then the equations (\ref{rownanienaznikanie2}) and (\ref{rownanienaznikaniepi}) reduce to: 
       \begin{equation} 
 \lambda_0=g4\sqrt{2}\ ,\qquad  \lambda_\pi=-g4\sqrt{2}\ .
       \end{equation} 
With $g=\frac{3}{4}\sqrt{2}k$ the bosonic part of the Lagrangian reads:
       \begin{equation} \label{lagrs}
	S=\int d^5 x \sqrt{-g_5} (\frac{1}{2}R + 6 k^2)-  6 \int d^5 x\sqrt{-g_4}k  (\delta(y) - \delta(y-\pi r_c))\ .
       \end{equation}

\subsection{Randall-Sundrum model with gravitini masses on the branes \cite{Bagger:2002rw}}
For the second example let us assume the prepotential of the 
form: $P_A^{\;B}=g{\rm i}(\sigma_1)_A^{\;B}$ and $(Q_0)_A^{\;B}=(Q_\pi)_A^{\;B}=(\sigma_3)_A^{\;B}$. Let us allow only the even components of gravitini 
to own mass terms on the branes
     \begin{equation}
	 (M_{0,\pi})_A^{\;B}=\frac{1}{2}\alpha_{0,\pi}(\sigma_1)_A^{\;B}\ ,\quad (\bar{M}_{0,\pi})_A^{\;B}=\frac{1}{2}\alpha_{0,\pi}{\rm i}(\sigma_2)_A^{\;B}\ ,
       \end{equation}
	where $\alpha_{0}, \alpha_{\pi}\in {\bf R}$.
In this case the equation (\ref{rownanienaznikanie2}) takes the form:
        \begin{eqnarray} \label{rownanienaznikaniebagger}
    0=&\delta(y){\rm i}\bar{\Psi}_\mu^A\gamma^\mu\bigg[&g\alpha_{0}(\delta-\gamma_5\sigma_{3})_{A}^{\;B}+\epsilon^{-1}(y)g\gamma_{5}\sigma_{1}(\delta+\gamma_5\sigma_{3})_{A}^{\;B}\nonumber\\&&+\frac{1}{4\sqrt{2}}\lambda_0\left(2\delta +\epsilon(y)\alpha_{0}\gamma_{5}\sigma_{1}(\delta+\gamma_5\sigma_{3})+(\delta+\gamma_5\sigma_{3})\right)_{A}^{\;B}\bigg]\eta_B\ ,
  \end{eqnarray} 
       and the boundary condition (\ref{warunkibrzeg}) become:  
    \begin{equation} \label{warunkibrzegbagger}
  \epsilon^{-1}(y) \delta(y) \gamma_5(\delta+\gamma_5\sigma_3)_A^{\;B}\eta_B=\delta(y)\alpha_{0}\sigma_1(\delta-\gamma_{5}\sigma_3)_A^{\;B}\eta_B\ .
    \end{equation} 
	One can also calculate boundary condition for the gravitino $ \bar{\Psi}_\mu^A$:
  \begin{equation} \label{warunkibrzegbaggergrav}
  \epsilon^{-1}(y) \delta(y) \bar{\Psi}_\mu^A(\delta-\gamma_5\sigma_3)_A^{\;B}\gamma_5=-\delta(y)\alpha_{0}\bar{\Psi}_\mu^A(\delta+\gamma_{5}\sigma_3)(\sigma_1)_A^{\;B}\ .
    \end{equation} 
	Finally, we can rewrite the equation (\ref{rownanienaznikaniebagger})
as follows
	  \begin{equation} \label{rownanienaznikaniebaggerfinall}
    0=\delta(y){\rm i}(\bar{\Psi}_{+})_\mu^A\gamma^\mu\left[2g\alpha_{0}+\frac{1}{4\sqrt{2}}\lambda_0(1+3\alpha^{2}_{0}\epsilon(y)^{2})\right](\delta-\gamma_5\sigma_{3})_{A}^{\;B}(\eta_+)_B\ ,
  \end{equation} 
	where we have decomposed fermions into the even $(+)$ and odd $(-)$
 components:
	\begin{equation} \label{defpmspinors}
	(\Psi_{\pm})_\mu^A=\frac{1}{2}(\delta\pm\gamma_5\sigma_{3})^{A}_{\;B}\Psi_\mu^B\ ,\quad (\eta_{\pm})^A=\frac{1}{2}(\delta\pm\gamma_5\sigma_{3})^{A}_{\;B}\eta^B\ .
	\end{equation} 
The vanishing of this term implies the constraint
	\begin{equation} 
	\lambda_0=-4\sqrt{2}g\frac{2\alpha_{0}}{1+\alpha^{2}_{0}}\ ,
	\end{equation} 
where we have used $\epsilon^{2}(y)\delta(y)=\frac{1}{3}\delta(y)$.
 One can repeat the same procedure for the second brane to obtain
	\begin{equation} 
	\lambda_\pi=-4\sqrt{2}g\frac{2\alpha_{\pi}}{1+\alpha^{2}_{\pi}}\ .
	\end{equation} 
Taking $\alpha_{0}=-\alpha_{\pi}=-1$ we find RS brane tensions $\lambda_{0}=-\lambda_{\pi}=g4\sqrt{2}$.
 
  Let us solve Killing equation to be certain that there remains an unbroken supersymmetry: 
       \begin{equation} 
	\nabla_\alpha\eta^A-\frac{\sqrt{2}{\rm i}}{3}{\cal P}^{A}_{\;B}\gamma_\alpha\eta^B-\delta(y)\delta_\alpha^5(\omega_{0})^{A}_{\;B}\gamma_5\eta^B-\delta(y-\pi r_c)\delta_\alpha^5(\omega_\pi)^{A}_{\;B}\gamma_5\eta^B=0\ .	 
       \end{equation} 
       For the RS background we can write
       \begin{eqnarray} 
       	&&0=\partial_\mu\eta_\pm^A-\frac{1}{2}k\epsilon(y)\gamma_\mu\gamma_5\eta_\mp^A+\frac{1}{2}k(\sigma_1)^A_{\;B}\gamma_\mu\eta_\pm^B\ ,\label{killingbag4}\\&&0=\partial_5\eta_+^A+\frac{1}{2}k(\sigma_1)^A_{\;B}\gamma_5\eta_-^B\ ,\label{killingbag5p}\\&&0=\partial_5\eta_-^A+\frac{1}{2}k(\sigma_1)^A_{\;B}\gamma_5\eta_+^B-2(\delta(y)-\delta(y-\pi r_c))\epsilon^{-1}(y)\eta_-^A\ .\label{killingbag5m}
	\end{eqnarray} 
The equation (\ref{killingbag4}) is solved by $\eta_-^A=\epsilon(y)\gamma_5(\sigma_1)^A_{\;B}\eta_+^B$, where we have assumed that the Killing spinor doesn't
 depend of $x_\mu$. 
       One can easily find the solution
       \begin{equation} \label{solutkillingspin}
       \eta_+^A=e^{-\frac{1}{2}k|y|}\left(\begin{array}{c}\hat{\eta}_R\\-\hat{\eta}_L\end{array}\right)^A\ ,\qquad \eta_-^A=\epsilon(y)e^{-\frac{1}{2}k|y|}\left(\begin{array}{c}\hat{\eta}_L\\\hat{\eta}_R\end{array}\right)^A\ ,
       \end{equation} 
       where $\hat{\eta}$ is a four--dimensional Majorana spinor
in flat space. 

\subsection{Flipped RS model with gravitini mass terms on the branes}
The flipped version of a warped 5d supergravity
without brane masses for gravitini has been described in detail in \cite{Brax:2001xf} and 
\cite{Lalak:2002kx}. Here let us perform the flipping (twisting) in the general case with 
brane gravitini masses. 

	To flip boundary condition in the fermionic sector, let us change a 
sign of the ${\bf Z}_2$ operator at the brane $y=\pi r_c$  to $(Q_\pi)=-(\sigma_3)$ while leaving $(Q_0)$ unchanged.  Let us examine the 
brane action at $y=\pi r_c$. We assume gravitini mass terms that include 
only (locally on each brane) even components of the fields 
	 \begin{equation}\label{masyflipped}
	 (M_\pi)_A^{\;B}=\frac{1}{2}\alpha_{\pi}(\sigma_1)_A^{\;B}\ ,\quad (\bar{M}_{\pi})_A^{\;B}=-\frac{1}{2}\alpha_{\pi}{\rm i}(\sigma_2)_A^{\;B}\ .
         \end{equation}
	Then boundary conditions change to
      \begin{equation} \label{warunkibrzegbaggerflip}
  \epsilon^{-1}(y) \delta(y-\pi r_c) \gamma_5(\delta-\gamma_5\sigma_3)_A^{\;B}\eta_B=-\delta(y-\pi r_c)\alpha_{\pi}\sigma_1(\delta+\gamma_{5}\sigma_3)_A^{\;B}\eta_B\ ,
      \end{equation}   
	and the relevant version of the equation 
(\ref{rownanienaznikaniebaggerfinall}) reads:
	 \begin{equation} \label{rownanienaznikaniebaggerflip}
    0=\delta(y-\pi r_c){\rm i}(\bar{\Psi}_{-})_\mu^A\gamma^\mu\left[2g\alpha_{\pi}+\frac{1}{4\sqrt{2}}\lambda_\pi(1+3\alpha^{2}_{\pi}\epsilon(y)^{2})\right](\delta+\gamma_5\sigma_{3})_{A}^{\;B}(\eta_-)_B\ ,
  \end{equation} 
	where $(+)$ and $(-)$ 
components are defined as in (\ref{defpmspinors}). Notice that now $(-)$ 
are even and the $(+)$ odd.
	Vanishing of this term implies constraint
	\begin{equation} 
	\lambda_\pi=-4\sqrt{2}g\frac{2\alpha_{\pi}}{1+\alpha^{2}_{\pi}}\ ,
	\end{equation} 
	As in the previous case, for $\alpha_{0}=-\alpha_{\pi}=-1$ we obtain RS brane tensions $\lambda_{0}=-\lambda_{\pi}=g4\sqrt{2}$.

   Let us check whether supersymmetry is broken in the effective theory, due to the flipped boundary condition. Killing equation reads:
       \begin{eqnarray} 
       	&&0=\partial_\mu\eta_\pm^A-\frac{1}{2}k\epsilon(y)\gamma_\mu\gamma_5\eta_\mp^A+\frac{1}{2}k(\sigma_1)^A_{\;B}\gamma_\mu\eta_\pm^B\ ,\label{killingbag4flipp}\\&&0=\partial_5\eta_+^A+\frac{1}{2}k(\sigma_1)^A_{\;B}\gamma_5\eta_-^B+2\delta(y-\pi r_c)\epsilon^{-1}(y)\eta_+^A\ ,\label{killingbag5pflipp}\\&&0=\partial_5\eta_-^A+\frac{1}{2}k(\sigma_1)^A_{\;B}\gamma_5\eta_+^B-2\delta(y)\epsilon^{-1}(y)\eta_-^A\ .\label{killingbag5mflipp}
	\end{eqnarray} 
       Equation (\ref{killingbag4flipp}) can be solved by $\eta_-^A=\epsilon(y)\gamma_5(\sigma_1)^A_{\;B}\eta_+^B$, where we assume that Killing spinors don't depend of $x_\mu$. 
       One can easily find solutions
       \begin{equation} \label{solutkillingspinflip}
       \eta_+^A=\epsilon_\pi(y)e^{-\frac{1}{2}k|y|}\left(\begin{array}{c}\hat{\eta}_R\\-\hat{\eta}_L\end{array}\right)^A\ ,\qquad \eta_-^A=\epsilon_0(y)e^{-\frac{1}{2}k|y|}\left(\begin{array}{c}\hat{\eta}_L\\\hat{\eta}_R\end{array}\right)^A\ ,
       \end{equation} 
       where $\epsilon_{0}(y)=\epsilon(\frac{y}{2})$ and $\epsilon_{\pi}(y)=\epsilon(\frac{y+\pi r_c}{2})$ are flipped step functions. 
 
This is an interesting example, especially if one compares of to the situation in the flipped sugra without brane mass terms, see \cite{Lalak:2002kx} and the forthcoming chapter. The projection operators on both branes are orthogonal, hence one 
would naturally expect that from the four-dimensional point of view 
all the supersymmetries are broken down. And this is the case indeed in the absence of brane mass terms. However, when the brane masses for gravitini are present, they constitute delta-type sources in the equations of motion of the gravitini. These become the actual source of the boundary conditions which have to be imposed on the bulk fields. The compatibility with the projection operators 
comes from the assumption, that all the odd fields can be written down 
as the step-function distribution centered on a given brane times the field 
that is even and continuous across the brane. Then one can formally assign
the value zero to such an odd field, but this zero should be assigned to 
the step-function distribution and not to the even field that multiplies the 
distribution. This is a consistent approach to the $\epsilon$-algebra,
see the comments in the footnote \ref{foot1}.
 
Hence, in a setup with brane masses the supersymmetry is controlled by 
boundary conditions and not by the projectors only. The boundary conditions do not need to be orthogonal even if the brane projectors are. In particular,
within the relations imposed on brane parameters and the prepotential, the 
change induced by the change of the brane projectors can be absorbed into the 
modification of fermionic mass terms on the branes without the necessity to 
modify the bosonic sources (like brane tensions in the example above). 
This is why one can have orthogonal brane projections, unbroken $N=1$ 
supersymmetry and Randall--Sundrum background. One should note however 
that the existence of the Killing spinor is by no means guaranteed even in the extended setup discussed here.

The bottom line of the above discussion is that flipped and unflipped RS models with gravitini mass terms on the branes, describe actually the same four--dimensional physics. One can easily check that the following redefinitions 
\begin{eqnarray} \label{redefinitionuf}
&\eta^+\longrightarrow\xi^+=\epsilon_\pi(y)\eta^+\ ,\quad\eta^-\longrightarrow\xi^-=\epsilon_\pi^{-1}(y)\eta^-\ ,&\nonumber\\&\Psi_\alpha^+\longrightarrow\chi_\alpha^+=\epsilon_\pi(y)\Psi_\alpha^+\ ,\quad\Psi_\alpha^-\longrightarrow\chi_\alpha^-=\epsilon^{-1}_\pi(y)\Psi_\alpha^-\ ,&
\end{eqnarray} 
transform unflipped model into the flipped one ($\eta$ and $\Psi$ are fields of the unflipped model and  $\xi$, $\chi$ - of the flipped model). The modifications of the supersymmetry transformations, (\ref{omegapi}), in both models are strictly connected by the redefinition (\ref{redefinitionuf}). To see the precise relation between respective mass terms let us concentrate on the 
terms in the unflipped action that are proportional to $\delta(y-\pi r_{c})$:
\begin{eqnarray} \label{redefuf} 
&&\frac{1}{2}\Bar{\Psi}^A_\mu\gamma^{\mu\nu}\gamma^5\partial_5\Psi_{\nu A}-\frac{1}{2}\delta(y-\pi r_{c})\Bar{\Psi}^A_\mu\gamma^{\mu\nu}\sigma_1(\delta-\gamma_5\sigma_3)_A^{\;B}\Psi_{\nu B}\longrightarrow\nonumber\\&&\longrightarrow\quad\frac{1}{2}\Bar{\chi}^A_\mu\gamma^{\mu\nu}\gamma^5\partial_5\chi_{\nu A}-\frac{1}{2}\epsilon_\pi^{-2}(y)\delta(y-\pi r_{c})\Bar{\chi}^A_\mu\gamma^{\mu\nu}\sigma_1(\delta-\gamma_5\sigma_3)_A^{\;B}\chi_{\nu B}\nonumber\\&&\qquad\,+\frac{1}{2}\epsilon_\pi^{-1}(y)\delta(y-\pi r_{c})\Bar{\chi}^A_\mu\gamma^{\mu\nu}\gamma^5(\delta-\gamma_5\sigma_3)_A^{\;B}\chi_{\nu B}\nonumber\\&&\qquad\,-\frac{1}{2}\epsilon^{-1}_\pi(y)\delta(y-\pi r_{c})\Bar{\chi}^A_\mu\gamma^{\mu\nu}\gamma^5(\delta+\gamma_5\sigma_3)_A^{\;B}\chi_{\nu B}\ .
\end{eqnarray} 
Boundary conditions derived from the new action are the same as (\ref{warunkibrzegbaggerflip}). On the other hand we can use these conditions in the last two terms of (\ref{redefuf}) to restore canonical action of the flipped model with mass terms (\ref{masyflipped}).

\section{Super--bigravity}
To introduce flipped boundary condition in the FLP model \cite{Brax:2001xf}, \cite{Lalak:2002kx}, let us assume $(Q_0)_A^{\;B}=-(Q_\pi)_A^{\;B}=(\sigma_3)_A^{\;B}$ and the prepotential $P_A^{\;B}=g_R\epsilon(y){\rm i}(\sigma_3)_A^{\;B}+g_S{\rm i}(\sigma_1)_A^{\;B}$. For $(M_0)_{AB}=(M_\pi)_{AB}=0$ one finds:
      \begin{equation} 
 \lambda_0=g_R4\sqrt{2}\ ,\qquad  \lambda_\pi=g_R4\sqrt{2}\ .
       \end{equation} 
      For $k=\frac{2\sqrt{2}}{3}\sqrt{g_{R}^{2}+g_{S}^{2}}$ and $T=g_{R}/\sqrt{g_{R}^{2}+g_{S}^{2}}$ we obtain the bosonic part of the action:
       \begin{equation} \label{lagads}
	S= \int d^5 x \sqrt{-g_5} (\frac{1}{2}R + 6 k^2)-  6 \int d^5 x\sqrt{-g_4}k T (\delta(y) +\delta(y-\pi r_c))\ .
       \end{equation}
	 One should notice that the brane tensions have the same sign. As a consequence gravitational background has no flat 4d Minkowski foliation, and the consistent 
solution is that of $AdS_4$ branes:
        \begin{equation} 
        ds^{2}=a^{2}(y)\bar{g}_{\mu\nu}dx^{\mu}dx^{\nu}+dy^{2}\ ,
        \end{equation}  
        where 
        \begin{equation} 
        a(y)=\frac{\sqrt{-\bar{\Lambda}}}{k}\cosh\left(k|y|-\frac{k\pi r_{c}}{2}\right)\ ,
        \end{equation} 
        and $\bar{g}_{\mu\nu}dx^{\mu}dx^{\nu}=\exp(-2\sqrt{-\bar{\Lambda}}x_{3})(-dt^{2}+dx^{2}_{1}+dx_{2}^{2})+dx_{3}^{2}$ is the four dimensional $AdS$ metric. 

The radius of the fifth dimension is determined by brane tensions:
        \begin{equation} 
        k\pi r_{c}=\ln\left(\frac{1+T}{1-T}\right)\ .
        \end{equation} 
        Normalization $a(0)=1$ leads to the fine tuning relation $\bar{\Lambda}=(T^{2}-1)k^{2}<0$.
       Compactification of this model has been performed 
in \cite{Lalak:2002kx}. Five--dimensional vacuum spontaneously brakes all supersymmetries due to the flipped boundary condition in the fermionic sector. 

Let us try to restore $N=1$ supersymmetry  in super--bigravity.   
To this end let us assume the prepotential of the form: $P_A^{\;B}=g{\rm i}(\sigma_1)_A^{\;B}$ and $(Q_0)_A^{\;B}=(Q_\pi)_A^{\;B}=(\sigma_3)_A^{\;B}$. 
Gravitini masses on the branes read
     \begin{equation}
	 (M_{0,\pi})_A^{\;B}=\frac{1}{2}\alpha_{0,\pi}(\sigma_1)_A^{\;B}\ ,\quad (\bar{M}_{0,\pi})_A^{\;B}=\frac{1}{2}\alpha_{0,\pi}{\rm i}
(\sigma_2)_A^{\;B}\ .
       \end{equation}
Taking
      \begin{equation} 
	\alpha_0=-\frac{\cosh(k\pi r_{c}/2)\pm 1}{\sinh(k\pi r_{c}/2)}\ ,\quad	\alpha_\pi=-\frac{\cosh(k\pi r_{c}/2)\pm 1}{\sinh(k\pi r_{c}/2)},
      \end{equation} 
      we obtain bosonic action of the bigravity model (\ref{lagads}). Notice, that we have two possibilities for the gravitini masses: $\alpha_0=1/\alpha_\pi$ and $\alpha_0=\alpha_\pi$. One can check that the first one is related to the BFL case \cite{Brax:2001xf} (both give the same physics), hence we shall explore in detail 
the other one. For simplicity, let us assume 
$\alpha_0=\alpha_\pi=-\alpha$, where
        \begin{equation} 
	\alpha=\frac{\cosh(k\pi r_{c}/2)- 1}{\sinh(k\pi r_{c}/2)}\ .
      \end{equation} 
      Then the boundary conditions take the form
      \begin{eqnarray} 
	&&\epsilon^{-1}(y) \delta(y) \gamma_5\eta^A_-=\delta(y)\alpha(\sigma_1)^A_{\;B}\eta^B_+\ ,\label{warbrzadsbigrav1}\\	&&\epsilon^{-1}(y) \delta(y-\pi r_c) \gamma_5\eta^A_-=-\delta(y-\pi r_c)\alpha(\sigma_1)^A_{\;B}\eta^B_+\label{warbrzadsbigrav2}\ .
      \end{eqnarray} 
	
   Let us now check whether supersymmetry is broken in the effective theory. 
The Killing equations read:
       \begin{eqnarray} 
       	&&0=\bar{\nabla}_\mu\eta_\pm^A+\frac{1}{2}k\epsilon(y)\tanh\left(k|y|-\frac{k\pi r_{c}}{2}\right)\gamma_\mu\gamma_5\eta_\mp^A+\frac{1}{2}k(\sigma_1)^A_{\;B}\gamma_\mu\eta_\pm^B\ ,\label{killingbag4bigrav}\\&&0=\partial_5\eta_+^A+\frac{1}{2}k(\sigma_1)^A_{\;B}\gamma_5\eta_-^B\ ,\label{killingbag5bigrav}\\&&0=\partial_5\eta_-^A+\frac{1}{2}k(\sigma_1)^A_{\;B}\gamma_5\eta_+^B-2(\delta(y)-\delta(y-\pi r_c))\epsilon^{-1}(y)\eta_-^A\ ,\label{killingbag5mbigrav}
	\end{eqnarray} 
       where $\bar{\nabla}_\mu$ denotes covariant derivative with respect to the four--dimensional $AdS$ geometry. One can decompose Killing spinors as 
follows:
       \begin{equation} 
       \eta_+^A=\phi_+(y)\left(\begin{array}{c}\hat{\eta}_R\\-\hat{\eta}_L\end{array}\right)^A\ ,\qquad\eta_-^A=\phi_-(y)\epsilon(y)\left(\begin{array}{c}\hat{\eta}_L\\\hat{\eta}_R\end{array}\right)^A\ ,
       \end{equation}  
       where $\hat{\eta}$ denotes killing spinor in the $AdS_4$, which satisfies: 
      $
       \left(\bar{\nabla}_\mu-\frac{1}{2}\sqrt{-\bar{\Lambda}}\hat{\gamma}_\mu\right)\hat{\eta}=0\ .
      $
       
The equation (\ref{killingbag4bigrav}) can be solved by 
       \begin{equation} 
	 \phi_-(y)=-\tanh\left(k|y|/2-k\pi r_{c}/4\right)\phi_+(y)\ .
       \end{equation} 
       One can easily check that boundary conditions (\ref{warbrzadsbigrav1}), (\ref{warbrzadsbigrav2}) are indeed satisfied. 

The equations (\ref{killingbag5bigrav}) and (\ref{killingbag5mbigrav}) take 
the form:
       \begin{eqnarray} 
       &&0=\partial_5\phi_+-\frac{1}{2}k\epsilon(y)\tanh\left(\frac{k|y|}{2}-\frac{k\pi r_{c}}{4}\right)\phi_+\ ,\label{killingbag5bigravfinall}\\&&0=\partial_5\phi_--\frac{1}{2}k\epsilon(y)\coth\left(\frac{k|y|}{2}-\frac{k\pi r_{c}}{4}\right)\phi_-\ \label{killingbag5mbigravfinall}.
       \end{eqnarray}
       One finds the following solutions:
       \begin{equation} \label{solutkillingspinbigrav}
       \phi_+=N\cosh\left(k|y|/2-k\pi r_{c}/4\right)\ ,\qquad \phi_-=-N\sinh\left(k|y|/2-k\pi r_{c}/4\right)\ ,
       \end{equation} 
       where $N$ is a normalization constant.

       One can find the explicit form of the zero-mode of gravitini. The bulk equations of motion for the even and odd components read:
       \begin{equation}   
	 \gamma^{\mu\rho\nu}\nabla_{\rho}(\Psi^\mp_{\nu})^{A}-\gamma^{\mu\nu}\gamma^{5}\partial_{5}(\Psi^\pm_{\nu})^{A}-k\epsilon\tanh\left(k|y|-\frac{k\pi r_{c}}{2}\right)\gamma^{\mu\nu}\gamma^{\hat{5}}(\Psi^\pm_{\nu})^{A}-\frac{3}{2}k(\sigma_1)^{A}_{\;B}\gamma^{\mu\nu}(\Psi^\mp_{\nu})^{B}=0\ .
       \end{equation} 
       Boundary conditions take the same form as for the Killing spinor:
         \begin{eqnarray} 
	&&\epsilon^{-1}(y) \delta(y) \gamma_5(\Psi^-_\mu)^A=\delta(y)\alpha(\sigma_1)^A_{\;B}(\Psi^+_\mu)^B\ ,\label{warbrzadsbigravit1}\\	&&\epsilon^{-1}(y) \delta(y-\pi r_c) \gamma_5(\Psi^-_\mu)^A=-\delta(y-\pi r_c)\alpha(\sigma_1)^A_{\;B}(\Psi^+_\mu)^B\label{warbrzadsbigravit2}\ .
      \end{eqnarray} 
The solution follows:
	  \begin{eqnarray} 
	    &&(\Psi^+_\mu)^A=N\cosh\left(k|y|/2-k\pi r_{c}/4\right)\left(\begin{array}{c}(\hat{\psi}_{\mu})_R\\-(\hat{\psi}_\mu)_L\end{array}\right)^A\ ,\\&&(\Psi^-_\mu)^A=-N\epsilon(y)\sinh\left(k|y|/2-k\pi r_{c}/4\right)\left(\begin{array}{c}(\hat{\psi}_{\mu})_L\\(\hat{\psi}_\mu)_R\end{array}\right)^A\ ,
      \end{eqnarray} 
	 where $\hat{\psi}_\mu$ denotes massless 4d gravitino in $AdS_4$:
       $
  \gamma^{\mu\rho\nu}\nabla_{\rho}\hat{\psi}_{\nu}=-\sqrt{-\bar{\Lambda}}\gamma^{\mu\nu}\hat{\psi}_{\nu}.
        $

Hence, we have managed to construct the model, which has the supersymmetric 
vacuum state where distance between branes is fixed, and the bosonic gravitational sector looks from the four--dimensional perspective as the $(++)$ 
bigravity \cite{Kogan:1999wc},\cite{Kogan:2000vb}. Dimensional reduction of the fermionic sector shall lead to 
truly $N=1$ supersymmetric four--dimensional version of bigravity (super--bigravity). Despite the fact that the background value of the 4d cosmological constant is non--zero, this is not a bad starting point for realistic model building. First of all, one can tune the value of the cosmological constant to be arbitrarily close to zero (this corresponds to the interbrane distance going towards 
infinity), second - the vacuum is static, with fixed expectation value of the 
radion, third - the vacuum energy shall be modified by radiative corrections after including matter into the game. The situation is not that dramatically different from that of the flat RS model: there the flat foliation implies 
that the interbrane distance is undetermined and one needs to complicate the 
model to fix the radion and break supersymmetry, which induces further corrections to the vacuum energy. Although it is possible to built a (somewhat baroque)
model where all the contributions conspire to give a vanishing vacuum energy in 4d, the procedure leading to that goal may be equally well 
applied to the present bigravity background.  

Finally, let us notice, that the model constructed in this section 
may be interpreted as the deconstruction\footnote{For the general idea of deconstruction see \cite{Arkani-Hamed:2001ca},\cite{Hill:2000mu}.} of 5d supergravity. First of all,
it is immediate to extend the model to the multi--brane setup of $S^1/\Pi \, {\bf Z_2}$ as in \cite{Lalak:2002kx}. Then
one can check, that in the limit of infinite interbrane distance all wave functions of gravitons and gravitini become localized on the branes, hence one has 
a simple product of K $SO(1,3)$ gravities localized on the branes. When the 
expectation value of the radion becomes finite (the radion `condenses'),
the bosonic and fermionic modes combine to form a single set of zero modes,
corresponding to the diagonal subgroup of the product of K $SO(2,3)$  
gravities, since for finite vev of the radion 4d geometry is the $AdS_4$.  

\section{Summary}

In this work we have extended previous analysis of the bulk/brane 
supersymmetrizations involving non-zero brane mass terms of bulk fermions 
(gravitini), and twisting of boundary conditions.  Our results confirm 
these of the ref. \cite{Bagger:2002rw} at points where the papers overlap.
We have applied the results 
to the construction of new brane/bulk models that may be relevant for 
realistic model building. In particular, we have built a model with the 
Randall--Sundrum bosonic sector, orthogonal projection operators on the branes in the fermionic sector, and an unbroken $N=1$ supersymmetry. 
We have also constructed 5d super--bigravity which allows for a static 
vacuum with unbroken $N=1$ supersymmetry. This model, in its multi--brane version, may be viewed as the deconstruction of 5d supergravity.  

\vspace{.6cm} 
\noindent We thank Max Zucker for very helpful and enjoyable
discussions.   
Z.L. thanks Theory Division at CERN for hospitality.
\noindent This work  was partially supported  by the EC Contract
HPRN-CT-2000-00152 for years 2000-2004, by the Polish State Committee for Scientific Research grant KBN 2P03B 129 24, and by POLONIUM 2003.

\vspace{.6cm}

\small{
}

\end{document}